# Elastic properties of superconducting LiFeAs from first principles

I.R. Shein,* and A.L. Ivanovskii

*Institute of Solid State Chemistry, Ural Branch of the Russian Academy of Sciences, Eekaterinburg, GSP-145, 620990, Russia*

**Abstract**

The first-principles FLAPW-GGA calculations of the elastic properties of recently discovered superconducting LiFeAs are reported. The independent elastic constants ($C_{ij}$), bulk modulus, compressibility, and shear modulus are evaluated and discussed. Additionally, numerical estimates of the elastic parameters of the polycrystalline LiFeAs ceramics are performed for the first time.

* Corresponding author.
*E-mail address:* shein@ihim.uran.ru (I.R. Shein).

# 1. Introduction

The recent discovery of high-temperature superconductivity ($T_C \sim$ 26-56K) [1-5] in so-called Fe-As systems has opened a new avenue for high-$T_C$ materials researches besides cuprates. The most numerous groups of these superconductors (SCs) - so-called FeAs "1111" and "122" phases - are based on quaternary (oxyarsenides *Ln*FeAsO, where *Ln* = La, Ce … .Gd, Tb, Dy) or ternary (arsenides *A*Fe$_2$As$_2$, where *A* are Ca, Sr, Ba) compounds. The most remarkable features of the mentioned FeAs SCs are that these phases adopt a quasi-two-dimensional crystal structures with alternating …[FeAs]$^{\delta-}$/[*Ln*O]$^{\delta-}$… (…[FeAs]$^{\delta-}$/[*A*]$^{\delta-}$…) blocks or atomic sheets; the non-doped parent compounds (*Ln*FeAsO or *A*Fe$_2$As$_2$) are non-superconducting and commonly exhibit a collinear spin-density waves, and the superconductivity emerges by their hole or electron doping, see review [6].

Quite recently the unique intrinsic superconductor LiFeAs ($T_C \sim$ 18K, which does not show any spin-density wave behavior but exhibits superconductivity at ambient pressure without chemical doping) was discovered, and some properties for this material are studied [7-14]. Besides the high scientific interest, today some advanced applications for these FeAs materials are discussed - for example, in very high magnetic field applications, as materials for thermoelectric cooling modules in the liquid nitrogen temperature range etc, see review [6]. Thus, the elastic properties of FeAs-based SC's are of great importance for their material science in view of future technological applications. Unlike some "1111" and "122" FeAs phases, for which the first estimations of their elastic parameters are performed recently [15,16], any data about mechanical behavior of LiFeAs are absent.

In view of these circumstances, in this paper we report the first-principles analysis of elastic properties of the recently synthesized superconductor LiFeAs.

# 2. Calculation details

The recently synthesized LiFeAs adopts tetragonal crystal structure [7-9] which contains [FeAs] blocks, based on edge-sharing tetrahedral {FeAs$_4$}, alternating with the atomic sheets of Li ions. Our calculations were performed using the FLAPW method (code WIEN2k) [17] with the GGA approximation of the exchange-correlation potential [18]. The radii of the atomic *muffin-tin* spheres were 2.2 a.u. (Fe), 2.3 a.u. (Li) and 2.0 a.u. (As). The set of plane waves $K_{max}$ was determined as $R_{MT}K_{max}$= 7.0. The convergence criterion was 0.01 mRy for the total energy and 0.01 $e$ for the charges.

## 3. Results and discussion

As the first step, six independent elastic constants ($C_{ij}$; namely $C_{11}$, $C_{12}$, $C_{13}$, $C_{33}$, $C_{44}$ and $C_{66}$) for the tetragonal LiFeAs were evaluated by calculating the stress tensors on different deformations applied to the equilibrium lattice of the tetragonal unit cell, whereupon the dependence between the resulting energy change and the deformation was determined, Table 1.

Firstly, for LiFeAs, all these elastic constants were positive and satisfied the well-known Born's criteria for tetragonal crystals: $C_{11} > 0$, $C_{33} > 0$, $C_{44} > 0$, $C_{66} > 0$, $(C_{11} - C_{12}) > 0$, $(C_{11} + C_{33} - 2C_{13}) > 0$ and $\{2(C_{11} + C_{12}) + C_{33} + 4C_{13}\} > 0$.

Secondly, the calculated elastic constants allowed us to obtain the macroscopic mechanical parameters for LiFeAs, such as bulk modulus ($B$) and shear modulus ($G$) – for example, using the Voigt (V) [19] approximation, as:

$$B_V = 1/9\{2(C_{11} + C_{12}) + C_{33} + 4C_{13}\};$$
$$G_V = 1/30(M + 3C_{11} - 3C_{12} + 12C_{44} + 6C_{66}\};$$

where $C^2 = (C_{11} + C_{12})C_{33} - 2C_{13}^2$ and $M = C_{11} + C_{12} + 2C_{33} - 4C_{13}$. The results obtained are presented in Table 1. From these data we see that for LiFeAs phase $B_V > G_V$; this means that a parameter limiting the mechanical stability of this material is the shear modulus. That is obvious enough due to the layered structure of this phase.

Next, as LiFeAs species are usually prepared and investigated as polycrystalline ceramics (see [7-9]) in the form of aggregated mixtures of micro-crystallites with a random orientation, it is useful to estimate the corresponding parameters for these polycrystalline materials from the elastic constants of the single crystals.

For this purpose we also calculated monocrystalline bulk modulus ($B$) and shear modulus ($G$) in Reuss approximation (R: $B_R$ and $G_R$, see Table 1) [20], and then we utilized the Voigt-Reuss-Hill (VRH) approximation. In this approach, according to Hill [21], the Voigt and Reuss averages are limits and the actual effective moduli for polycrystals could be approximated by the arithmetic mean of these two limits. Then, one can calculate [15,16] the averaged compressibility ($\beta_{VRH} = 1/B_{VRH}$), Young modulus ($Y_{VRH}$), and from $B_{VRH}$, $G_{VRH}$ and $Y_{VRH}$ it is possible to evaluate the Poisson's ratio ($v$). All these parameters are listed in Table 3. Certainly, all these estimations were performed in the limit of zero porosity of LiFeAs ceramics.

From our results we see that the bulk modulus for LiFeAs is rather small (< 100 GPa) and became less than, for example, the bulk moduli for other known superconducting species such as $MgB_2$, $MgCNi_3$, YBCO and $YNi_2B_2C$ for which $B$ vary from 115 to 200 GPa [22-26]. Thus, as compared with these SCs, LiFeAs is soft material – as well as the related $SrFe_2As_2$ and LaFeAsO phases, Table 2.

Finally, according to the criterion [27], a material is brittle if the $B/G$ ratio is less than 1.75. In our case, for LiFeAs this value is 1.59, respectively. This means that LiFeAs belongs to the relatively brittle materials. Besides, the values of the Poisson ratio ($v$) are minimal for covalent materials ($v = 0.1$) and grow essentially for ionic species [28]. In our case, the value of $v$ for LiFeAs is about 0.24, *i.e.* a considerable ionic contribution in intra-atomic bonding takes place.

## 4. Conclusions

In summary, by means of first-principles FLAPW-GGA total energy calculations we have predicted for the first time the elastic properties for mono- and polycrystalline LiFeAs. Our analysis showed that this phase is a mechanically stable anisotropic material. The parameter limiting its mechanical stability is the shear modulus. LiFeAs is relatively soft material ($B$ < 100 GPa) with high compressibility and will behave as brittle system.

**Acknowledgements**

This work was supported by FRBR, Grant No 09-03-00946-a

**Table 1.** The calculated elastic constants ($C_{ij}$, in GPa), bulk modulus ($B$, in GPa) and shear modulus ($G$, in GPa) for tetragonal monocrystalline LiFeAs.

| | |
|---|---|
| $C_{11}$ | 176 |
| $C_{12}$ | 55 |
| $C_{13}$ | 62 |
| $C_{33}$ | 131 |
| $C_{44}$ | 56 |
| $C_{66}$ | 72 |
| $B_V$ ($B_R$) * | 93.4 (92.1) |
| $G_V$ ($G_R$) * | 57.1 (58.9) |

* in Voigt (Reuss) approximation

**Table 2**. Calculated values of some elastic parameters for polycrystalline LiFeAs ceramics as obtained in the Voigt-Reuss-Hill approximation: bulk modulus ($B_{VRH}$, in GPa), compressibility ($\beta_{VRH}$, in GPa$^{-1}$), shear modulus ($G_{VRH}$, in GPa), Young modulus ($Y_{VRH}$, in GPa) and Poisson's ratio ($\nu$) – in comparison with SrFe$_2$As$_2$ and LaFeAsO [15,16].

| oxypnictide | $B_{VRH}$ | $\beta_{VRH}$ | $G_{VRH}$ | $Y_{VRH}$ | $\nu$ |
|---|---|---|---|---|---|
| LiFeAs | 92.8 | 0.01078 | 58.0 | 144.0 | 0.241 |
| SrFe$_2$As$_2$ [16] | 61.7 | 0.01621 | 2.3 | 6.8 | 0.482 |
| LaFeAsO [15] | 97.9 | 0.01022 | 56.2 | 141.5 | 0.259 |